\documentstyle[psfig,mnextra]{mn}

\def\gsim{~\rlap{$>$}{\lower 1.0ex\hbox{$\sim$}}}

\title{Diffuse X-ray Emission from Late-Type Galaxy Haloes}
\author[A.~J.~Benson, R.~G.~Bower, C.~S.~Frenk and S.~D.~M.~White]
{A.~J.~Benson$^{1,3}$, R.~G.~ Bower$^{1,4}$, C.~S.~Frenk$^{1,5}$ and S.~D.~M.~White$^{2,6}$ \\
1. Physics Department, University of Durham, South Road, Science Laboratories, Durham DH1 3LE, England. \\
2. Max Plank Institute f\"{u}r Astrophysik, Karl-Schwarzchild Strasse 1, D-85740, Garching, Germany. \\
3. E-mail: A.J.Benson@dur.ac.uk \\
4. E-mail: R.G.Bower@dur.ac.uk \\
5. E-mail: C.S.Frenk@dur.ac.uk \\
6. E-mail: swhite@mpa-garching.mpg.de
}

\begin{document}

\maketitle

\begin{abstract}
Current theories of galaxy formation predict that spiral galaxies are
embedded in a reservoir of hot gas. This gas is able to cool onto the
galaxy replenishing cold gas that is consumed by star
formation. Estimates of the X-ray luminosity emitted in the cooling
region suggest a bolometric luminosity of order $10\times10^{41}\hbox{
ergs s}^{-1}$ in massive systems. We have used ROSAT PSPC data to
search for extended X-ray emission from the haloes of three nearby,
massive, late-type galaxies: NGC~2841, NGC~4594 and NGC~5529. We infer
95\% upper limits on the bolometric X-ray luminosities of the haloes
of NGC~2841, NGC~4594 and NGC~5529 of 0.4, 1.2 and 3.8 $\times 10^{41}
\hbox{ ergs s}^{-1}$ respectively. Thus, the true luminosity lies well
below the straightforward theoretical prediction. We discuss this
discrepancy and suggest a number of ways in which the theoretical
model might be brought into agreement with the observational
results. A possible solution is that the gravitational potentials of
these galaxies' dark matter haloes are weaker than assumed in the
current model. Alternatively, the present day accretion may be
substantially less than is required on average to build the disk over
the Hubble time. Our results are, however, based on only three
galaxies, none of which is ideal for this kind of study. A larger data
set is required to explore this important problem further.
\end{abstract}


\section{Introduction}

In models of galaxy formation by hierarchical clustering, the disks of
spiral galaxies form by late accretion of gas which cools from an
extended reservoir around the galaxy. In these models disks are still
expected to be growing at the present day. Such models have been very
successful in accounting for properties of galaxies at optical and
infrared wavelengths (see, for example, Kauffmann et al. 1993, Cole et
al. 1994, Baugh, Cole \& Frenk 1996, Kauffmann et al. 1999). The
proposed disk formation mechanism should produce a signature at X-ray
wavelengths which, in principle, is detectable with the present
generation of satellites.

At the virial temperature of galaxy halos, the dominant cooling
mechanism is X-ray Bremsstrahlung. If the cooling rate is significant,
this flux may be visible as a component of diffuse X-ray emission
extending well beyond the galaxy's optical radius. In this paper, we
set out to test this disk formation model by looking for X-ray
emission from the hypothesized reservoir of cooling gas around three
large, late-type galaxies, NGC~2841, NGC~4594 and NGC~5529.

If we assume that the mass of the present disk has been built up by a
constant rate of accretion over the age of the universe ($t_0$), a
simple estimate of the X-ray luminosity can be made from the binding
energy of the disk.
\begin{equation}
L_X \approx {{1\over2}M_{\mathrm disk} v_{\mathrm esc}^2\over t_0}
\label{eq:be}
\end{equation}
where $M_{\mathrm disk}$ is the present mass of the disk, $v_{\mathrm
esc} = V_{\mathrm c} \left[ 2 \ln (r_{\mathrm vir}/r_{\mathrm disk}) +
2 \right] ^{1/2}$ is the velocity required by material at the edge of
the disk to escape the halo (assuming a flat rotation curve) and
$V_{\mathrm c}$ is the measured circular velocity of the
disk. Adopting parameters suitable for a large spiral galaxy such as
NGC~2841 (ie., $M_{\mathrm disk}=5\times10^{10} M_\odot$, and
$V_{\mathrm c}=317\,\hbox{km/s}$), this suggests that the halo should
emit a bolometric luminosity of at least $\sim 8\times10^{41}\,\hbox{
ergs s}^{-1}$\footnote{Unless otherwise stated we have assumed
${\mathrm H}_0 = 50 \hbox{ km s}^{-1} \hbox{Mpc}^{-1}$, $\Omega _0 =
1$, $\Lambda _0 = 0$, $\Omega_{\mathrm b}=0.06$ and $\sigma _8 =
0.67$}. If supernovae and other feedback processes are effective in
heating halo gas then even more energy must be radiated to assemble
the observed disk. Note that eqn. (\ref{eq:be}) predicts a minimum
$L_{\rm X} \propto V_{\rm c}^5$, and that the temperature of the
cooling gas should scale as the halo virial temperature, $T_{\rm X}
\propto V_{\rm c}^2$. Thus, it is important to search for emission in
the most massive systems available.

The picture of hierarchical galaxy formation developed by White \&
Frenk (1991, WF91; see also Kauffmann et al. 1993, Cole et al., 1994)
predicts halo X-ray luminosities more directly by computing the
cooling rate of gas as a function of time. At the present epoch, such
models (which are tuned to reproduce the observed optical properties
of galaxies) predict an X-ray luminosity of about $10 \times 10^{41}
\hbox{ ergs s}^{-1}$ for galaxies like NGC~2841.

\begin{table*}
\begin{center}
\caption{Observed properties of the galaxies. Shown for each galaxy
are its heliocentric radial velocity, distance, $W_{20}$ (the width of
the 21cm line at 20\% of its peak intensity), $N_{\rm H}$ (the column
density of neutral hydrogen in the Milky Way in the direction of the
galaxy), $r_{\rm optical}$ (the 25 B-magnitudes per square arcsecond
isophotal radius of each galaxy measured along the major axis),
$M_{\rm B}$ (the absolute B-band magnitude) and morphological type.}
\label{tb:obsprops}
\begin{tabular}{lrrrrrrr}
\hline
 & \textbf{Heliocentric radial} & & $\mathbf{W_{20}}$ & $\mathbf{N_H}$ & $\mathbf{r_{\mathbf{optical}}}$ & & \textbf{Hubble} \\
\textbf{Name} & \textbf{velocity (km/s)}$^a$ & \textbf{Distance (Mpc)} & \textbf{(km s$^{\mathbf{-1}}$)}$^d$ & \textbf{($\mathbf{10^{20}}$ cm$^{\mathbf{-2}}$)}$^e$ & \textbf{(arcmin)}$^d$ & ${\mathbf M_{\mathrm B}}$ & \textbf{type}\\
\hline
NGC 2841 & $638 \pm 3.0$ & $13.8 \pm 5.2^b$ & $674 \pm 4$ & 1.45 & $4.1 \pm 0.04$ & $-20.6 \pm 0.2$ & Sb \\
NGC 4594 & $1091 \pm 5.1$ & $24.54 \pm 2.5^c$ & $753 \pm 7$ & 3.67 & $4.4 \pm 0.04$ & $-23.0 \pm 0.2$ & Sa \\
NGC 5529 & $2885 \pm 5.1$ & $57.72\pm 2.5^c$ & $592 \pm 5$ & 1.09 & $3.3 \pm 0.07$ & $-21.1 \pm 0.2$ & Sc \\
\hline
\end{tabular}
\begin{flushleft}
$^a$ From the NASA Extragalactic Database (NED). \\
$^b$ de Vaucoulers (1979) \\
$^c$ From heliocentric radial velocity plus peculiar velocity correction assuming ${\mathrm H}_0 = 50 \hbox{ km s}^{-1}\hbox{ Mpc}^{-1}$.\\
$^d$ From the RC3 catalogue (available from NED), corrected for inclination. \\
$^e$ Dickey \& Lockman (1990). \\
\end{flushleft}
\end{center}
\end{table*}

To date, studies of X-ray emission from nearby galaxies (e.g. Read,
Ponman \& Strickland, 1997) have concentrated on X-ray emission on
scales comparable to the optical image of the galaxy. While extended
emission has been detected in some groups of galaxies (e.g. Mulchaey
et al. 1996, Trinchieri, Kim \& Fabbiano 1994, Trinchieri, Fabbiano \&
Kim 1997), this emission is thought to be associated with the group
potential. Our aim here is to test for the equivalent emission in {\it
isolated} spiral galaxies, which we would identify with a cooling flow
supplying the galaxy disk with gas. A few studies have reported
diffuse emission from isolated elliptical galaxies. However, this has
been interpreted either as resulting from the expulsion of gas from
the galaxy (e.g. Read \& Ponman 1998, Mathews \& Brighenti 1998) or as
a relic of a collapsed group (Ponman et al. 1994). Furthermore, since
in hierarchical models of galaxy formation ellipticals are not
necessarily expected to be accreting gas at the present day, the
existence or otherwise of a diffuse X-ray component associated with
them does not provide a strong test of such theories. For this reason,
we have restricted our study to isolated spiral galaxies.

Previous X-ray observations of spirals (see, for example, Fabbiano \&
Juda 1997, Burstein et al. 1997, Pellegrini \& Ciotti 1998, Brighenti
\& Matthews 1999) do not place strong constraints on the models we aim
to test. These studies have focussed on measuring X-ray fluxes in
regions comparable to the optical radius of the galaxy, where the
intensity is strongest. X-ray emission from galactic stellar sources
and uncertainties in the spatial X-ray surface brightness profile make
it difficult to determine what fraction, if any, of the observed X-ray
flux could be due to a low surface brightness cooling
flow. 

Since the cooling gas must, at some point, settle into the galaxy's
disk, some emission is expected from this region as the gas cools from
the virial temperature of the halo to much lower temperatures. For
example, in the case of NGC~4594 models of the type considered here
predict a luminosity comparable to that measured by Fabbiano \& Juda
(1997), $8.9 \pm 1.4 \times 10^{40}$ ergs/s in the 0.1-2.4keV band
within the optical radius of the galaxy. However, likely contamination
by stellar sources complicates the interpretation of this comparison.

In order to test the models in a manner free from uncertainties of the
type just described it is essential to look for low surface brightness
emission at large distances from the optical disk of the galaxy. It is
also necessary to study the optically brightest galaxies since the
expected X-ray luminosity depends so strongly on the circular velocity
of the galaxy's halo. For this reason, we have chosen for this study
nearby galaxies with particularly high circular velocities. Most
previous work has studied relatively low mass galaxies from which only
relatively weak emission is predicted.

If diffuse X-ray emission from gas in the haloes of isolated spiral
galaxies proves undetectable, this need not imply that the mass of
halo gas is small since the gas may simply be too diffuse to emit
X-rays efficiently. However, it would then be unable to cool and flow
to the centre of the halo. We are thus testing whether the disks of
spiral galaxies are being built up by the accretion of halo gas at the
present day. The alternative is that spiral disks were assembled at
higher redshift and evolved as essentially closed systems thereafter.

The structure of our paper is as follows. In \S2, we search for
large-scale diffuse emission from massive spiral galaxies. We are not
able to make any convincing detection; the upper limits we establish
are almost an order of magnitude below the predictions of the WF91
model. In \S3 we discuss the uncertainties in the models and describe
how they might be modified to agree with the data. The prospects for
future investigations are outlined in \S4.
 
\section{Observations}
\label{obssec}

\subsection{Target Selection}

The primary aim of this paper is to search for diffuse X-ray emission
from nearby isolated spiral galaxies. The galaxy must be nearby for
the signal to be detectable. Furthermore, the galaxy must have a high
circular velocity in order that the energy released by the cooling gas
be large. A high circular velocity also favours detectability because
it implies a high characteristic temperature for the emission and so
less absorption by neutral hydrogen in the Milky Way. In addition the
efficiency of supernovae feedback is expected to decline with
increasing circular velocity.

We created a suitable target list by searching for all spiral galaxies
with recession velocities below 3000 $\hbox{km s}^{-1}$ and with HI
20\% line widths greater than 580 $\hbox{km s}^{-1}$. Of the sixteen
galaxies satisfying these criteria three have been observed by the
ROSAT PSPC: NGC 2841, NGC 4594 (the Sombrero galaxy) and NGC 5529. The
PSPC data for these fields were obtained from the Leicester Database
and Archive Service (LEDAS).

\begin{itemize}
\item NGC~2841. The Sb galaxy NGC~2841 is the nearest of our target
galaxies. It has a flat rotation curve extending to the limit of HI
observations, and is an archetype of galaxies with massive dark matter
haloes (Kent, 1987). Its star formation properties have been studied
extensively by Young et al.\ (1996) using both $H\alpha$ and far
infra-red indicators.

\item NGC~4594. Also known as M104 or the ``Sombrero Galaxy,'' this is
a highly luminous Sa seen almost edge on. Since this galaxy is bulge
dominated, diffuse X-ray emission similar to that seen around E or S0
galaxies (Forman, Jones \& Tucker 1985) might be expected. However,
X-ray observations by the \emph{Einstein} satellite revealed spectral
properties and core X-ray colours typical of less massive spiral
galaxies (Kim, Fabbiano \& Trinchieri 1992).

\item NGC~5529. This is an Sc galaxy seen almost edge-on. Because of
its relatively large distance, this galaxy is likely to provide weaker
constraints than the previous two.
\end{itemize}

Distances, line widths, optical radii, and other relevant properties
of these galaxies are listed in Table \ref{tb:obsprops}. The distances
to two of them (NGC 4594 and NGC 5529) had to be determined from their
observed heliocentric radial velocity; a model of the local peculiar
velocity field (Branchini et al. 1999) was used to correct the
observed velocity to the true Hubble velocity using an iterative
procedure. Table \ref{tb:vctemp} lists the circular velocity of each
galaxy as calculated from its HI line-width and also the virial
temperature of the halo estimated, as in WF91.

\begin{table}
\begin{center}
\caption{Measured rotation velocity and predicted gas temperature for the
galaxies in our sample.} 
\label{tb:vctemp}
\begin{tabular}{lcr}
\hline
\textbf{Name} & $\mathbf{V_{\mathrm c}}$ (km s$^{-1}$)$^a$ & $\mathbf{T_{vir}}$ \textbf{(keV)}$^b$ \\
\hline
NGC 2841 & 317 $\pm$ 2 & 0.287 $\pm$ 0.004 \\
NGC 4594 & 358 $\pm$ 4 & 0.366 $\pm$ 0.009 \\
NGC 5529 & 277 $\pm$ 3 & 0.219 $\pm$ 0.005 \\
\hline
\end{tabular}
\begin{flushleft}
$^a$ Calculated from HI line widths using the expression given in Tully \&
Fouqu\'e (1985) \\
$^b$ As defined by White \& Frenk (1991)
\end{flushleft}
\end{center}
\end{table}

\subsection{Data Reduction}

Ideally, we would like to compare the X-ray flux from an annulus
centered on the galaxy (but excluding the optically bright region to
avoid contamination from stellar emission) with similar annuli
centred on blank fields. This is not possible, however, because
different ROSAT fields have different particle backgrounds. We have
therefore adopted the procedure of first subtracting the detected
events from a large radius annulus in the outer part of the PSPC
detector as described in Snowden et al. (1994). Background corrected
`on source' and `off source' fields may then be compared to determine
the flux from the diffuse halo of the target galaxies. It is important
to note that the background subtraction algorithm of Snowden et al.
tends to produce negative flux in the off-source images because of the
different point-source detection thresholds in the centre and outer
parts of the ROSAT field. Our analysis takes this effect into account.
 
\begin{figure*}
\begin{center}
\begin{tabular}{cc}
\psfig{file=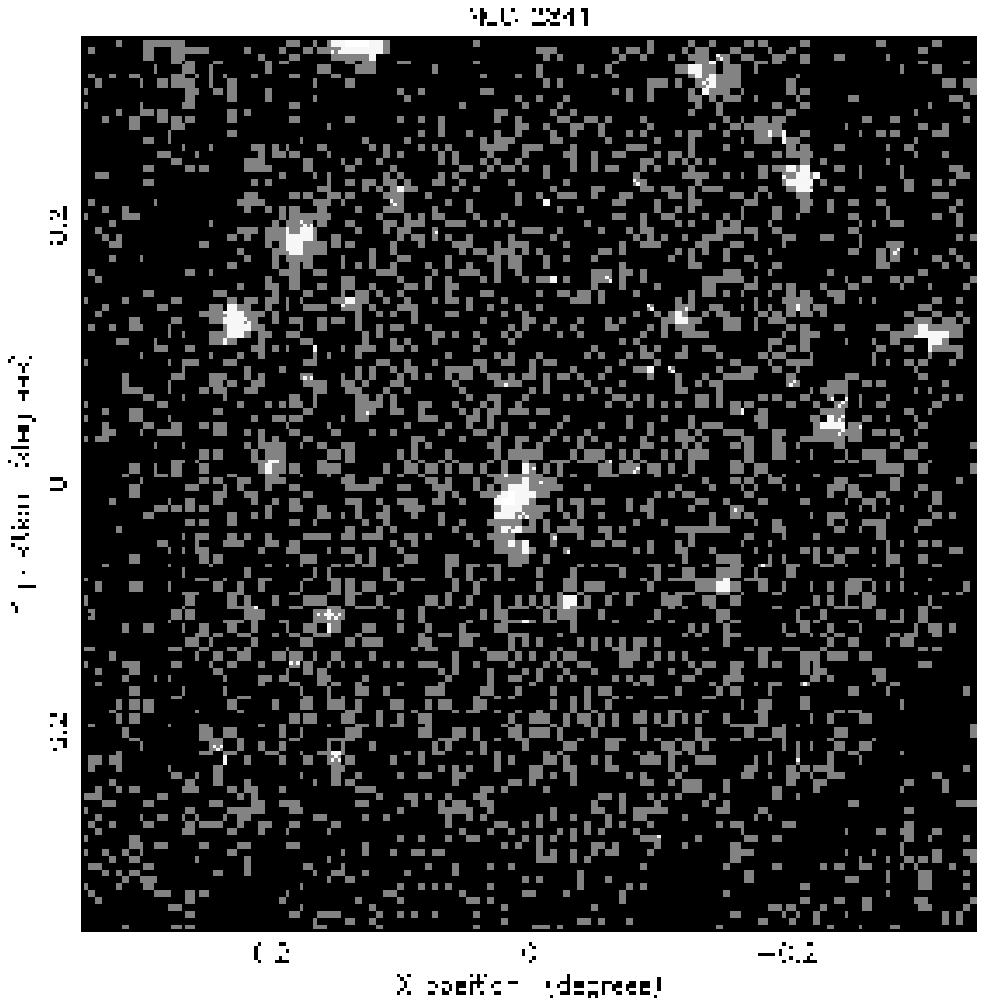,width=74mm,bbllx=55mm,bblly=90mm,bburx=160mm,bbury=190mm,clip=,angle=270} & \psfig{file=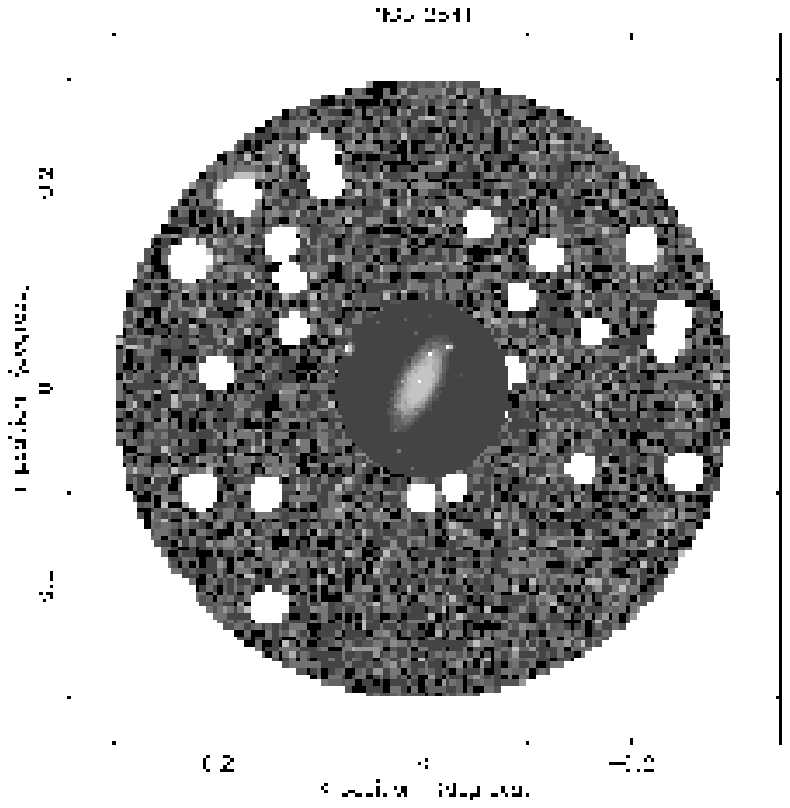,width=80mm,bbllx=60mm,bblly=100mm,bburx=150mm,bbury=190mm,clip=} \\
\psfig{file=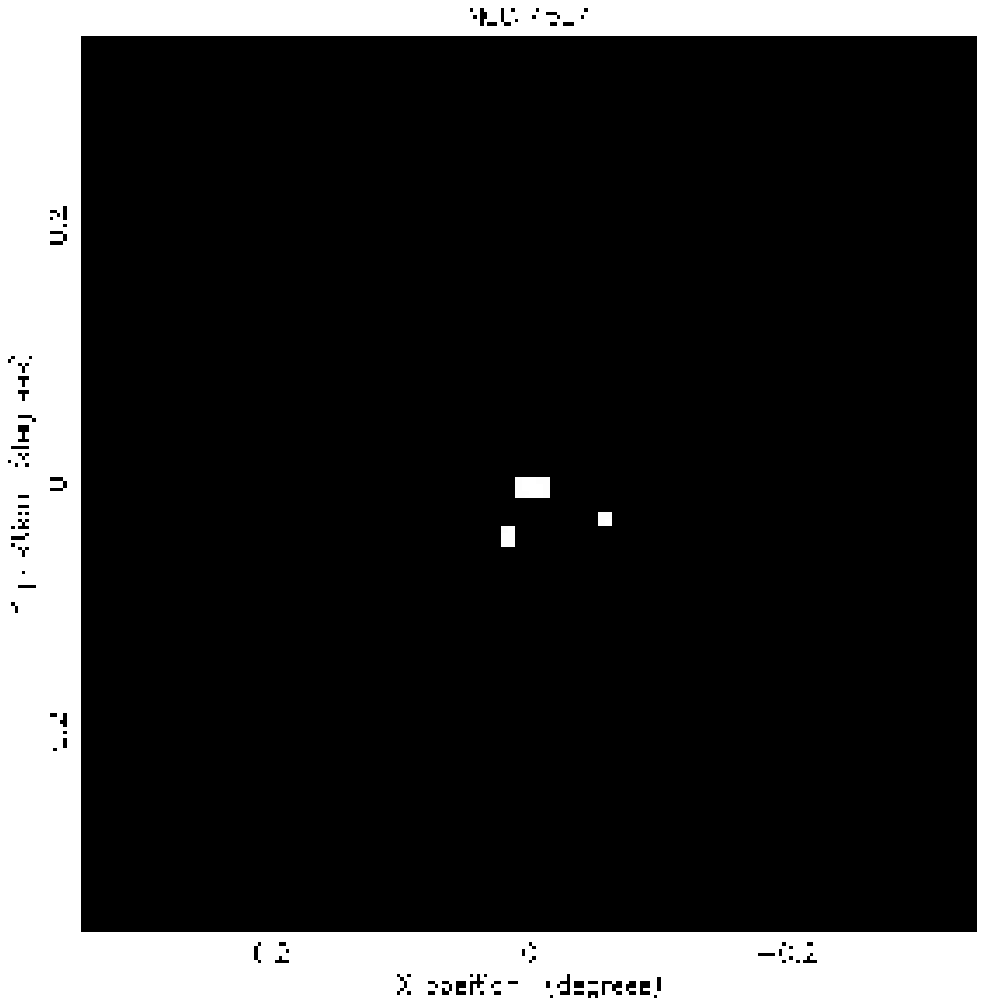,width=74mm,bbllx=55mm,bblly=90mm,bburx=160mm,bbury=190mm,clip=,angle=270} & \psfig{file=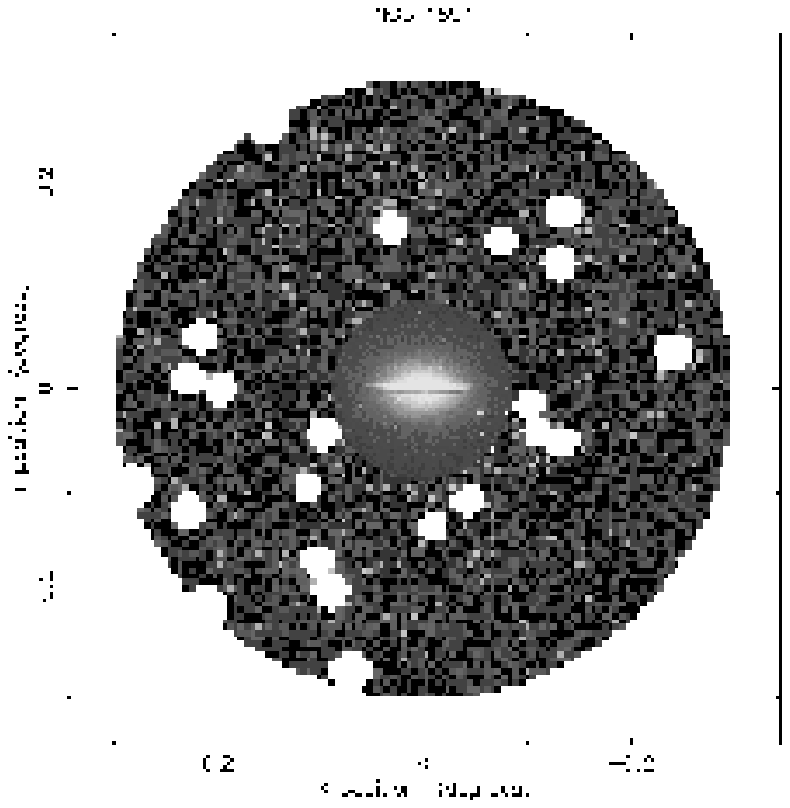,width=80mm,bbllx=60mm,bblly=100mm,bburx=150mm,bbury=190mm,clip=} \\
\psfig{file=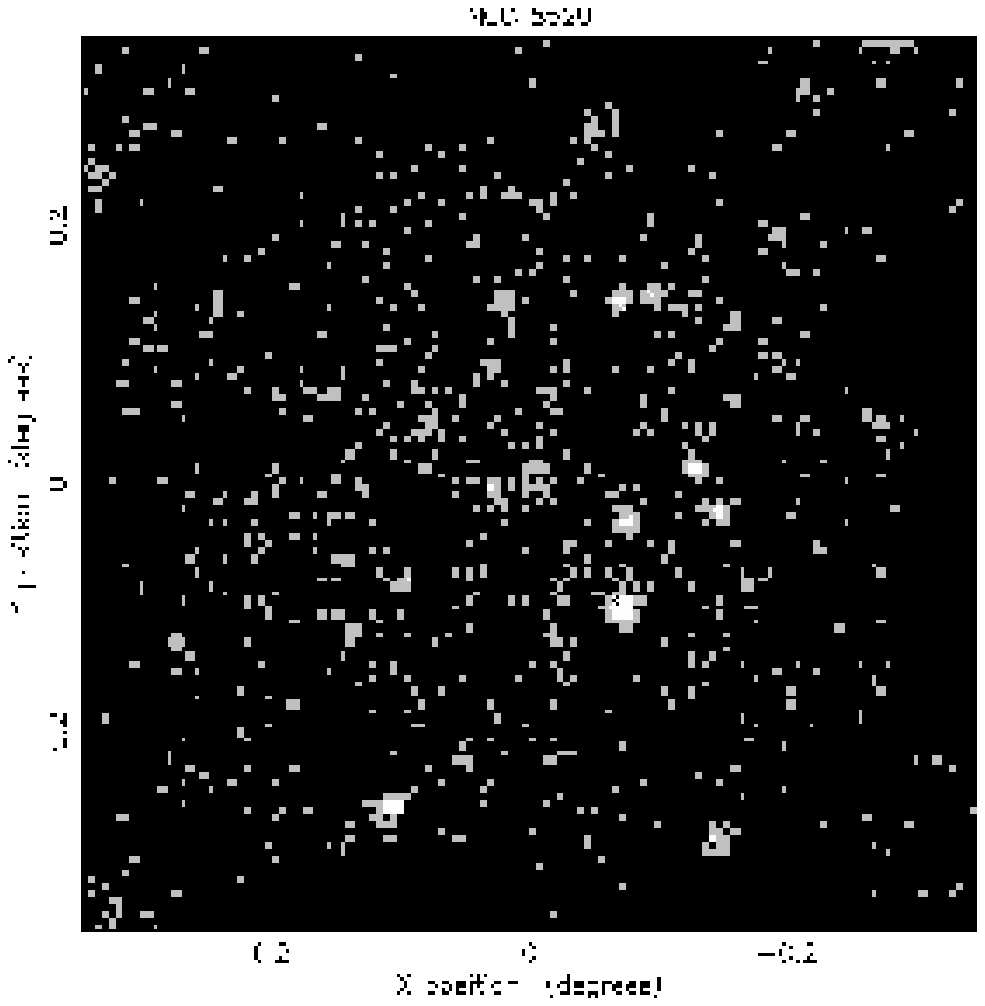,width=74mm,bbllx=55mm,bblly=90mm,bburx=160mm,bbury=190mm,clip=,angle=270} & \psfig{file=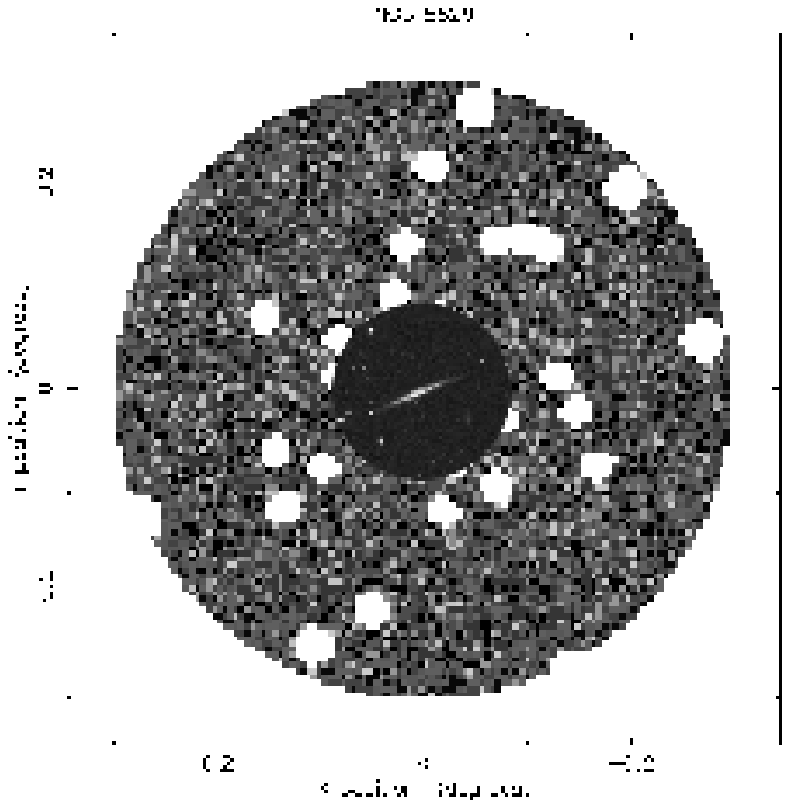,width=80mm,bbllx=60mm,bblly=100mm,bburx=150mm,bbury=190mm,clip=}
\end{tabular}
\caption{Source images for each galaxy in the sample. The left hand column
shows the raw image and the right hand column the image after background
subtraction, point source removal and selection of the region from 5 to 18
arcminutes. An optical image of each galaxy is superimposed on the X-ray
image to give an appreciation of the size of the halo region.}
\label{fig:sources}
\end{center}
\end{figure*}

The analysis of the data was carried out using the Asterix
package. The procedure for analysing each image was as follows.
Firstly, point sources were detected down to a $3 \sigma$ threshold
and excluded by cutting out a circle centred on each source. The
radius of this circle was set equal to the radius of the PSF enclosing
95\% of the energy at the position of each source. For each image a
background annulus of inner radius $0.5^{\circ}$ and outer radius
$0.7^{\circ}$ was then used to estimate the vignetting corrected
background count-rate, and to create two background-subtracted spectra
from source-centred annuli with inner radius of 5 arcminutes and outer
radii of 9 and 18 arcminutes. Choosing a fixed angular radius ensures
that the blank field analysis is the same for each image. The outer
radius was chosen to include the entire inner ring of the PSPC. This
is comparable to the predicted X-ray emitting region of the closer of
our target galaxies (see \S3). The $9'$ ring is matched to the X-ray
emitting region of the more distant galaxy NGC~5529. The innermost
aperture was excluded from the analysis in order to avoid
contamination by the diffuse and unresolved stellar X-ray emission of
the galaxy itself (cf. Read, et al.\ 1997, for example). The X-ray
image of each galaxy is compared with the optical image in
Figure~\ref{fig:sources}.

The resulting spectra were analysed with the XSpec package to determine
the source flux using the \emph{mekal}
X-ray emission and \emph{wabs} absorption models. For each
halo we adopted a virial temperature of 0.2keV. This is slightly below
the virial temperatures estimated in Table~\ref{tb:vctemp}, in order
to obtain conservative estimates of the bolometric luminosity. The
hot halo gas is assumed to have a metallicity of $0.3 Z_{\odot}$
(where $Z_{\odot}$ is the Solar metallicity). Neutral absorption
column densities varied between the fields and were taken from Dickey
\& Lockman (1990). The only remaining free parameter in these models
is the overall normalization. This procedure appropriately weights the
contributions of the different energy channels and corrects for the
different neutral hydrogen cross-sections of the fields.

\begin{table*}
\begin{center}
\caption{X-ray fluxes and bolometric luminosities for the haloes of
three spiral galaxies. The fluxes and luminosities are determined for
two annuli, with inner radius of 5 arcminutes and outer radii of 9
(ring 1) and 18 (ring 2) arcminutes. The first column gives the name
of the galaxy.  The flux corrected for blank field offsets is given in
the following two columns, and the final two columns list the
bolometric luminosity of each halo corrected for blank field offsets.}
\label{obs:ann}
\begin{tabular}{lcccc}
\hline
 & \textbf{Bg. Corr.} & \textbf{Bg.\ Corr.} & \textbf{Bolometric} & \textbf{Bolometric} \\
 & \textbf{Flux (ring 1)} & \textbf{Flux (ring 2)} & $L_{\mathrm X}$ \textbf{(ring 1)} & $L_{\mathrm X}$ \textbf{(ring2)} \\
\textbf{Galaxy} & ($10^{-14}$ ergs/cm$^{2}$/s) & ($10^{-14}$ ergs/cm$^{2}$/s) & ($10^{41}$ ergs/s) & ($10^{41}$ ergs/s) \\
\hline
NGC~2841 & $5.62 \pm 10.33$ & $22.90 \pm 19.87$ & $0.03 \pm 0.06$ & $0.16 \pm 0.13$ \\
NGC~4594 & $17.96 \pm 11.33$ & $14.21 \pm 22.40$ & $0.44 \pm 0.28$ & $0.34 \pm 0.54$ \\
NGC~5529 & $5.91 \pm 10.03$ & $-2.28 \pm 19.22$ & $0.78 \pm 1.34$ & $-0.29 \pm 2.48$ \\
\hline
\end{tabular}
\end{center}
\end{table*}

\begin{table}
\begin{center}
\caption{The total counts per second in six blank fields. Background
and point sources were removed in each case and an annulus from 5 to 9
arcminutes (ring 1) and from 5 to 18 arcminutes (ring 2) was
selected.}
\label{tb:bf2}
\begin{tabular}{lcc}
\hline
\textbf{Field} & \textbf{Ring1} & \textbf{Ring 2} \\
\hline
rp201558 & $-0.017 \pm 0.007$ & $-0.063 \pm 0.015$ \\
rp700255 & $-0.004 \pm 0.006$ & $-0.040 \pm 0.013$ \\
rp700319 & $-0.021 \pm 0.005$ & $-0.024 \pm 0.011$ \\
rp700376 & $-0.007 \pm 0.005$ & $-0.003 \pm 0.011$ \\
rp700387 & $+0.007 \pm 0.004$ & $-0.013 \pm 0.008$ \\
rp700389 & $-0.005 \pm 0.006$ & $-0.023 \pm 0.013$ \\
\hline
\end{tabular}
\end{center}
\end{table}

The fitted spectrum is then used to calculate the absorption corrected
flux from each halo. These fluxes, shown in Table \ref{obs:ann}, are
within the range 0.2 to 2.0 keV in order to minimise dependence on the
assumed temperature. These are converted to pseudo-bolometric luminosities
(0.001 to 10.0 keV) using the conservative temperature of 0.2~keV and the
distances given in Table~\ref{tb:obsprops}.

For comparison, six off-source fields (Table \ref{tb:bf2}) with point
source targets were selected at random from the ROSAT International
X-ray/Optical Survey (RIXOS) (Mittaz et al. 1998). As we have
previously noted, the analysis procedure tends to produce a small
negative count-rate when applied to the off-source fields. This is due
(at least in part) to the different flux limits at which point sources
are removed from the target and background annuli. The mean luminosity
deficit in the off-source fields was added to the measured
luminosities of the target galaxies to correct for this
effect. Alternatively one could add the off-source field count rates
to the on-source fields and only then fit a model spectrum to the
resulting counts to determine a corrected luminosity. We investigated
performing the analysis in both ways and found that each procedure
gives the same result to within 10\%. Since the off-source count rates
vary from field to field it is possible that the true correction
required for each target field is somewhat different from the mean
correction from the off-source fields. We therefore measure the
variance in the off-source field correction and add this quantity in
quadrature to the errors on the fluxes measured in our target
fields. The variations from field to field are our dominant source of
uncertainty.

The luminosities of the target galaxies are presented in Table
\ref{obs:ann}. For each galaxy we expect a luminosity around $10
\times 10^{41}\hbox{ ergs s}^{-1}$. However none of the galaxies is
detected at a 3$\sigma$ confidence level in either of the annuli
considered. Instead, our data sets 95\% upper confidence limits of
$0.37, 1.2$ and $3.8 \times 10^{41}\hbox{ ergs s}^{-1}$ in the larger
aperture for NGC~2841, 4594 and 5529 respectively.  For NGC~5529, the
relatively weak upper limit is due to the galaxy's larger
distance. For the two more nearby galaxies, the upper limit lies an
order of magnitude below the initial theoretical prediction.

\subsection{Uncertainties in the Observations}

In \S\ref{sec:discuss}, we consider possible modifications to the
model of WF91 that might alleviate the discrepancy with our
data. First, we take a closer look at the assumptions we made in
interpreting the observational data. There are a number of uncertain
quantities that could affect our conclusions. Throughout the analysis
we have consistently assumed conservative values for these quantities
(i.e. those that would minimize the conflict between model and data.)
It is nevertheless instructive to consider how these choices might
affect our results.

The dominant source of uncertainty in our analysis is the process of
background subtraction. Although an accurate estimation of the
background from a mosaic of surrounding pointings would be preferable,
our off-source fields indicate the magnitude of the uncertainty. Even
using the `worst case' field (rp201558), the predicted and observed
count-rates still disagree by a factor of 2--3. 

Another important source of uncertainty is the extrapolation from
ROSAT count-rate to bolometric flux.  At the low temperatures
appropriate to galactic halo gas, neutral hydrogen absorption in our
own Galaxy has a strong effect. If our adopted values for the neutral
hydrogen column density were inaccurate, this could have a substantial
effect on our results. However these galaxies are at high Galactic
latitude and we find that increasing the assumed column density by a
factor of 3 over the values given by Dickey \& Lockman (1990) only
changes the inferred bolometric luminosity by 40\%.

The effects of background correction and absorption alone cannot
reconcile the observed and predicted luminosities. Altering the
assumed temperature of the halo plasma, however, is a more promising
alternative. This sensitivity arises because, at the relevant
temperatures, most of the emission lies outside the 0.2--2.0 keV band
of the observations, forcing us to extrapolate the model fit to lower
energies. For example, at a temperature of 0.2~keV, 90\% of the
bolometric luminosity is emitted in the ROSAT 0.2--2.0~keV band, while
at 0.1~keV, only 75\% of the emission lies in this band. At even lower
temperatures, the effect becomes even more pronounced. Applying a
lower temperature model to the observed spectra, we find that a
decrease in temperature by a factor of 4.0 (for NGC~2841) to 4.5 (for
NGC~4594) is required to bring the observations into line with the
predicted luminosity. It is not clear, however, how these lower
temperatures can be accommodated within the theoretical model. We now
turn to a detailed examination of this model.

\section{Discussion}
\label{sec:discuss}

The model presented by WF91 predicts that the X-ray luminosity due to
gas cooling within a dark matter halo should be:

\begin{equation}
L_{\mathrm X} = \dot{M}_{\mathrm cool}(r_{\mathrm cool}) V_{\mathrm c}^2
	\ln\left(r_{\mathrm cool}\over r_{\mathrm optical}\right),
\label{eq:simplecool}
\end{equation}

\noindent where $\dot{M}_{\mathrm cool}(r_{\mathrm cool})$ is the rate
at which mass is cooling in the dark matter halo, $r_{\mathrm cool}$
is the cooling radius (which grows with time), and $r_{\rm optical}$
is the optical radius of the galaxy, which we take to be the extent of
the 25 B-magnitudes per square arcsecond isophote along the major
axis. This model assumes that the halo has an isothermal potential
with circular velocity $V_{\mathrm c}$. As shown in the Appendix, the
model can be readily extended to the case where the gravitational
potential follows the functional form favoured by Navarro, Frenk \&
White (1996, NFW).

We have used the formulae and parameter choices in the Appendix to
predict the bolometric X-ray luminosity for each of the three galaxies
(using the circular velocities of Table \ref{tb:vctemp}), as well as
the region in which cooling is expected to occur.  Table
\ref{tb:finalpredrcool} lists the cooling radii and luminosities of
the haloes predicted by WF91's standard model and by a model with an
NFW potential. In column~5, we list the luminosity that is expected
within the 5-$18'$ annulus used in the observational
measurements. This geometric correction is made under the assumption
of an isothermal flow in quasi-hydrostatic equilibrium as described in
the Appendix. The figure given includes a correction for the flux from
the target galaxy that is expected in the outer background annulus by
extrapolating beyond the cooling radius with an $r^{-2}$ density
profile.  The corrected flux is about 50\% and 70\% of the total for
NGC~2841 and NGC~4594 respectively. The larger correction for NGC~2841
arises because the emission from the target almost fills the PSPC
field of view.  The geometric correction for NGC~5529 is greater
because a large proportion of the luminosity is projected onto the
optical image of the galaxy. In the case of this galaxy, the large
observational uncertainties mean that the emission model is
only weakly constrained.

For both NGC~2841 and NGC~4594 the bolometric X-ray luminosities
predicted by the model are strongly inconsistent with the
observations. The WF91 model predicts cooling rates of 7.4 and $9.0
M_\odot \hbox{yr}^{-1}$ for NGC~2841 and 4594 respectively. Since the
mass cooling rate is proportional to the X-ray luminosity in these
theories, the simplest interpretation is that the cooling rates in
these galaxies are also much less than those predicted by WF91. If we
assume that the geometric factors would be unchanged by reducing the
cooling rate, we arrive at upper limits to the present-day accretion
rate of 0.5 and $1.44 M_\odot \hbox{yr}^{-1}$.

Further insight can be gained by considering the total mass of gas
within the cooling radius and the mass of the galaxy disk. In the case
of NGC~2841 the disk mass can be determined by fitting the galaxy
rotation curve. Kent (1987) derives a disk mass of $9.3\times10^{10}
M_\odot$ allowing a free-fit of disk and isothermal halo
parameters. Adopting an NFW model for the halo, Navarro (1998, N98)
derives a best-fit disc mass of $5 \times 10^{10} M_{\odot}$; a
significantly smaller disk mass would be inconsistent with the blue
luminosity of the disk (Young et al., 1996).  The WF91 model
prediction for the mass of gas within $r_{\mathrm cool}$, which is
assumed to have formed the disk, is $9.8 \times 10^{10}
M_{\odot}$. Thus the model allows enough gas to cool to make the
observed disk. Consequently X-ray observations imply that the
present-day accretion rate does not significantly contribute to the
total disk mass. Clearly, this is an important result, since (if these
galaxies are representative) it implies that the masses of spiral
disks have remained almost constant over recent look-back
times. Finally, we can compare the gas accretion rate with the star
formation rate in NGC~2841 of $0.8 M_{\odot}\hbox{yr}^{-1}$ inferred
from H$\alpha$ emission (Young et al.\ 1996). Thus, although our data
exclude an infall rate sufficient to build the disk over a Hubble
time, they may allow enough infall to replenish the gas being used up
by star formation.

Before reaching wide-ranging conclusions, however, we must ask whether
it is possible to modify the emission model to bring it more into line
with the observational data.  Below we consider possible modifications
to the halo model that might reduce the discrepancy with the
observational results.

\begin{description}
\item[{\bf Feedback.}] We have chosen galaxies with particularly deep
potential wells in order to minimise the uncertainties arising from
supernovae feedback effects on our predictions. According to the
models of Kauffmann et al. (1993) and Cole et al. (1994) feedback in
$\sim 300$ km/s halos reduces the predicted star formation rate by
only $\sim 5\%$. Any gas ejected from the galaxy by feedback should
begin cooling once more, and may actually increase the X-ray
luminosity above our theoretical predictions. Feedback in these models
cannot, therefore, reconcile theory and observations. However, in
other models of galaxy formation, such as that of Nulsen \& Fabian
(1995, 1997) and Wu, Nulsen \& Fabian (1999), in which feedback in
earlier generations of dark matter halos alters the gas density
profile (thereby reducing the cooling rate), the expected X-ray
luminosities of spiral halos are likely to be substantially less than
in the models we have tested. Although these models do reduce the
luminosity of X-ray halos to levels consistent with our data, the
cooling rate at late times appears much too small to allow the growth
of disks as large and as massive as those observed. Further work is
needed to investigate this point further.

\item[{\bf Cosmological Parameters.}]  The predicted X-ray
luminosities have a relatively weak dependence on the values of
the cosmological parameters (here we adopted ${\mathrm H}_0 = 50 \hbox{
km s}^{-1} \hbox{Mpc}^{-1}$, $\Omega _0 = 1$, $\Lambda _0 = 0$,
$\Omega_{\mathrm b}=0.06$ and $\sigma _8 = 0.67$). The strongest
dependence is on the baryon fraction since the cooling rate is
proportional to the square of the density of the X-ray emitting
plasma. Using $\Omega_{\mathrm b} = 0.04$ (a 50\% reduction) reduces
the luminosities by a factor $\sim 2$, but at the expense of lowering
the mass cooling rate also by a factor $\sim 2$. Reasonable changes in
cosmological parameters are unable to lower the predicted X-ray
luminosity by more than a factor $\sim 3$. In particular, lowering
$\Omega_0$ tends to increase the cooling rate (and consequently the
X-ray luminosity) because the baryon fraction of the halo is then
increased.

\item[{\bf X-ray Emission Temperature.}] An assumption of the model is
that the X-ray emission occurs at a characteristic temperature close
to that of the galactic halo. If the plasma had a multiphase
structure, then the emission weighted temperature could be
lower. However, a 25\% decrease in temperature reduces the expected
flux by only 20\%. Much larger changes in temperature are required in
order to account for the discrepancy.

\item[{\bf Spatial Distribution of X-ray Emission.}] Because the
cooling radius estimated for NGC~2841 is so large, the correction for
the spatial distribution of the X-ray emission is important. In our
standard model, approximately 60\% of the total luminosity is emitted
within the 5--18 arcminute annulus which we observe. If, for example,
we considered a very different spatial distribution for the luminosity
(whilst keeping the total luminosity the same), such as one in which
the X-ray emissivity remains constant within the cooling radius, then
the fraction of the total luminosity emitted within the annulus would
drop to around 30\%. The luminosity (corrected for geometric effects)
listed in Table~\ref{tb:finalpredrcool} would therefore be reduced by
a factor of 2. These corrections are much less important for NGC~4594
where the cooling radius is better matched to the PSPC field of view.

\item[{\bf Halo Circular Velocity.}]  The luminosity of the flow is
based on the premise that the high circular velocities of the disks of
these galaxies reflect the circular velocities of the halo that
confines the hot plasma. Recently, several authors (Zaritsky et al.\
1997, N98) have suggested that high circular velocity galaxies could
have formed in significantly cooler haloes. N98 has analyzed the
rotation curve of NGC~2841, and found a best fitting model with
$V_{200}=167^{+134}_{-13}$ km s$^{-1}$ (where $V_{200}$ is the
circular velocity at the virial radius). Using the corresponding
potential in the calculation of the X-ray luminosity we predict a
substantially reduced X-ray emission of $0.9 \times 10^{41}$ ergs
s$^{-1}$ (including geometric factors), in much better agreement with
the observational results. The ratio of the predicted value to the
observational upper limit is 0.62. Unfortunately such a model
underpredicts the disk mass inferred by N98 by a factor of
4. Increasing the gas fraction to remedy this problem restores the
predicted X-ray luminosity to an unacceptably high value.

\item[{\bf Metal abundance.}]  We have assumed that the hot gas in the
haloes has a metal abundance similar to that measured in galaxy
clusters, namely $Z = 0.3 {\mathrm Z}_{\odot}$. Using $Z = 0.5
{\mathrm Z}_{\odot}$ increases the predicted X-ray luminosity by
approximately 30\%, since gas with a higher metal content can cool
more efficiently. Reducing the abundance to $0.1 {\mathrm Z}_{\odot}$
would, for the same reason, reduce the predicted luminosities (and gas
accretion rates), by a factor of $\approx 3$, but would also reduce
disk masses by the same factor.

\item[{\bf Remaining gas fraction.}] The predicted luminosity also
depends upon the value of $f_{\rm g}$, the fraction of the baryonic
material in the halo that is in gaseous form at the time when the halo
forms. This parameter is unlikely to be much less than the value of
unity that we have assumed unless a significant fraction of the
baryons in galaxies are in some dark form. Baugh et al. (1998) find
that$f_{\rm g} > 0.7$ for halos of circular velocity comparable to
those of the galaxies considered here. If $f_{\rm g}$ were lowered
enough so that the predicted X-ray luminosity agreed with the observed
upper limits, then the gas cooling rate would be significantly lower,
making it impossible to build up a disk of the observed mass by the
present. Therefore, although we do not know the exact value of $f_{\rm
g}$, a lower value cannot reconcile the model with both the observed
X-ray luminosity and the inferred disk mass.

\item[{\bf Multiphase flow.}] We have assumed that gas flows into the
galaxy at the centre of the halo at a fixed temperature and only cools
once it has settled in the galaxy disk. However, it is conceivable
that the gas could instead cool ``in-situ'' (i.e. at the location in
the dark matter halo where it originated) and then fall into the
galaxy as cold clouds (Thomas, Nulsen \& Fabian 1987). If this were
indeed the case, then the bolometric X-ray luminosity emitted by the
cooling gas would be $2.5 \dot{M}_{\rm cool}V_{\rm C}^2$ (assuming
that the gas cools at constant pressure). For typical values of
$r_{cool}$ and $r_{\rm optical}$ this is about half the luminosity
predicted by eqn. (\ref{eq:simplecool}). Furthermore, this luminosity
would be emitted at a lower mean temperature than the virial
temperature of the halo. Using the cooling flow model of Mushotzky \&
Szymkowiak (1988) we estimate that, for gas cooling from 0.2 keV, 11\%
of the bolometric luminosity would be emitted in the 0.2-2.0 keV band
after accounting for absorption. This is approximately half of the
24\% that is emitted in this band for gas at a fixed temperature of
0.2 keV. Thus, in the multiphase model the predicted flux in the
0.2-2.0 keV band is lower by a factor of approximately 4 compared to
our standard model. These models are thus much less discrepant with
the observational upper limits.

It should be noted, however, that even in this scenario the infalling
gas must still release both its thermal and gravitational energy
somewhere. The discussion above accounts only for the thermal
component, and this model only succeeds if the gravitational component
is radiated outside the ROSAT energy band.  This not readily
accomplished: if the cold clouds fall ballistically towards the
centre, the gravitational energy would be radiated as post-shock
X-rays when they collide with the galaxy's gas disk. It does not help
if the cold clouds are coupled to the hot gas by magnetic fields since
although they could flow in more slowly by transferring their
gravitational energy to the hot component, this energy will ultimately
be radiated in the X-ray band.

Support for a multiphase model may be provided by the high-velocity
clouds recently detected by Blitz et al. (1999) if these are indeed
infalling onto the Milky Way. However, Blitz et al. estimate that the
present day infall rate of such clouds onto the Milky Way is around
$0.8 M_{\odot}$ yr$^{-1}$, and this is significantly lower than the
rate predicted by the models we consider ($\sim$ 5--6 $M_\odot$
yr$^{-1}$). Thus, unless the observed clouds represent only a small
fraction of the total cloud population, the simple multiphase model
that we have considered seems to conflict with the inferred infall
rate of cold gas onto the Milky Way.
\end{description}

\begin{table*}
\begin{center}
\caption{Predictions from our models, assuming the standard cosmology
and halo gas with $Z = 0.3 {\mathrm Z}_{\odot}$. Columns 3 and 4 list
the total model luminosity and cooling radius in arcmin; column~5
gives the model luminosity after correction for the geometry of the
source and background annuli; column~6 lists the ratio of the observed
luminosity to the model luminosity after correction for the geometry.}
\label{tb:finalpredrcool}
\begin{tabular}{llccccccc}
\hline
 & &  Model &$R_{\rm cool}$& Geo. corrected & Ratio of obs. 95\%\\
Target & Halo model & $L_{\mathrm X}/10^{41}$ ergs/s & (arcmin)& Model $L_{\mathrm X}/10^{41}$ ergs/s & upper limit to model\\
\hline 
NGC2841 & Isothermal & $10.4 \pm 0.10$ & $35.1\pm13.7$& $5.5 \pm 0.06$ & $<0.07$ \\
 & NFW & $ 7.6 \pm 0.74$ & $48.0\pm18.1$& $4.5 \pm 0.44$ & $<0.09$ \\
NGC4594 & Isothermal & $ 10.8 \pm 0.18$& $18.7\pm1.9$ & $7.6 \pm 0.13$ & $<0.16$ \\
 & NFW & $ 9.5 \pm 0.63$ & $23.0\pm2.4$ & $6.9 \pm 0.46$ & $<0.17$ \\
NGC5529 & Isothermal & $3.4 \pm 0.03$ & $9.4\pm0.4$ & $1.3 \pm 0.01$ & $<2.9$ \\
 & NFW & $ 2.4 \pm 0.10$ & $12.5\pm0.5$ & $0.9 \pm 0.04$ & $<4.2$ \\
\hline
\end{tabular}
\end{center}
\end{table*}

\section{Conclusions}

We have searched for large-scale diffuse X-ray emission from three
nearby, high rotation velocity spiral galaxies. In each case, we find
that the observed emission is almost {\it an order of magnitude} below
that required to build the disk at a constant rate within the age of
the universe. The accretion rate is similarly discordant with
theoretical estimates of the cooling rate of gas in an isothermal or
NFW-profile halo.

Although extremely suggestive, our results are not definitive. The
reasons for this are: \begin{itemize}
\item Strong constraints are derived only for the two nearer
galaxies. We may, by chance, have selected two systems that have
anomalous cooling rates. This is plausible for the bulge dominated
NGC~4594 galaxy (`the Sombrero'). NGC~2841 appears to be more typical
of spiral galaxies, but its proximity makes accurate background
subtraction difficult, and our results are dependent on the profile
assumed for the X-ray emission. By contrast, the third galaxy meeting
our selection criteria is too distant for strong limits on the X-ray
flux to be set.  We note, however, that the X-ray limits established
for these galaxies are compatible with the emission that is thought to
occur in the halo of our own galaxy (eg., Moore \& Davis, 1994,
Wolfire, et al.\ 1995).

\item There are a number of parameters that have been assumed in order
to predict the X-ray flux from the properties of the halo. No
parameter, or combination of parameters, can simultaneously give a low
X-ray luminosity and build the observed disk. To reconcile the
predictions with the observed luminosities we must assume that either:
(a) disks were built at high redshift; or, (b) accretion onto disks
occurs without radiating the binding energy of the gas at X-ray
wavelengths (e.g. in a multiphase flow).
\end{itemize}

Our results encourage further work. The current upper limits are set
by our ability to subtract the background count-rate in the
images. This is exacerbated in the ROSAT data because the flux level
at which individual point sources may be subtracted varies across the
image on scales comparable to the spiral galaxy haloes. Future work
requires large-scale imaging with a detector that has a near-uniform
detection threshold, such as the EPIC camera aboard the XMM satellite
(Lumb et al. 1996). Such a survey would need to target a greater
fraction of the nearby galaxies, setting firmer limits on the diffuse
X-ray emission and eliminating variations in galaxy properties as a
major source of uncertainty.  If such work continues to fail to make
detections, then we may be forced to conclude that spiral disks are no
longer growing at the present epoch. This would agree with traditional
models of galaxy formation in which galaxies accrete their mass at
early times and evolve as more or less closed systems to the
present-day. Clearly, there is much to learn from future studies of
the X-ray haloes of spiral galaxies.

\section*{Acknowledgements}

The authors would like to acknowledge Shaun Cole and Julio Navarro for
providing rotation curve modelling code and best fit parameters for
NGC~2841 respectively. We thank Enzo Branchini for making available to
us his peculiar velocity field models prior to publication and Alan
Smith for his work on a pilot version of this project as part of his
undergraduate studies at the university of Durham. We have made use of
the NASA Extragalactic Database (NED) and the Leicester Database and
Archive Service (LEDAS). This project was carried out using computing
facilities supplied by the Starlink Project, and was supported by a
PPARC rolling grant for ``Extragalactic Astronomy and Cosmology at
Durham'' and by the European Community's TMR Network for Galaxy
Formation and Evolution. AJB and CSF acknowledge receipt of a PPARC
Studentship and Senior Research Fellowship respectively. CSF also
acknowledges a Leverhume Fellowship.

\appendix

\section{A Simple Model for the X-ray Emission from Cooling Gas}

Following WF91, we assume that cooling occurs within a static
potential, and that the cooling gas remains isothermal as it flows to
the centre. The assumption of isothermality allows considerable
simplification in the determination of the X-ray luminosity of the
flow since it is directly related to the drop in potential energy
(ie., the internal energy of the gas remains constant). Detailed,
fully self-consistent flow models, such as those of Bertschinger
(1989) and Abadi et al. (1999), show that these assumptions are
reasonable over the range of scales over which we wish to determine
the flow solution. The X-ray luminosity emitted by a unit mass of gas
as it moves within the potential, $\phi$, is

\begin{equation}
l_{\mathrm X} = -v_{\mathrm R} {\partial \phi \over \partial r} .
\label{eq:lumpervol}
\end{equation}

In models such as those of Kauffmann et al. (1993) and Cole et
al. (1994) gas is assumed to flow from the cooling radius towards the
centre of the halo. No mass is assumed to fall out of the flow until
it reaches the central galaxy. We find the total energy emitted by
unit mass of gas as it flows inwards and then integrate from
$r_{\mathrm cool}$ to $r_{\mathrm optical}$ to find the total energy
emitted by the mass of gas,

\begin{equation}
q = \int ^{r_{\mathrm cool}}_{r_{\mathrm optical}} -{l_{\mathrm X}(r) \over v_{\mathrm R}(r)} {\mathrm d}r.
\end{equation}

\noindent If we now make the assumption that mass flows to the centre
more quickly than the structure of the flow changes, then the structure
of the flow at a snapshot in time can be represented by the
trajectory of a single mass unit. Hence

\begin{equation}
L_{\mathrm X} = \dot{M}_{\mathrm cool}(r_{\mathrm cool}) \int ^{r_{\mathrm cool}}_{r_{\mathrm optical}} {V_{\mathrm c}^2(r) \over r} {\mathrm d}r .
\label{eq:appLX}
\end{equation}

Below we consider two possibilities for the structure of the dark
matter halo. In \S\ref{isocoolsec} we consider an isothermal halo and
in \S\ref{NFWcool} we consider the dark matter profile described by
NFW, in each case assuming that the gas initially traces the dark
matter.

\subsection{Cooling in an Isothermal Halo}

\label{isocoolsec}

This potential was considered by WF91.  The dark matter density
profile is $\rho (r) \propto r^{-2}$, where $r$ is radial distance in
the halo.  The cooling radius, $r_{\mathrm cool}$, in the halo is
defined as the radius at which the cooling time, $t_{\mathrm cool}$,
of the gas equals the age of the Universe, $t_0$. Ideally, we would
use the time since the halo last doubled in mass (Cole \& Lacey,
1993), but this information is not available for an individual
galaxy. Since mass cooling rates decline with time, this leads to a
conservative comparison with our observational results.

Assuming a spherically symmetric halo in which the gas is initially
distributed like the dark matter, the cooling time is given by

\begin{equation}
t_{\mathrm cool} (r) = {192 \pi \Omega _0 {\mathrm G m_p^2} r^2 \over 49 f_{\mathrm g} \Omega _{\mathrm b}
\Lambda ( \mu {\mathrm m_p} V_{\mathrm c}^2 / {\mathrm 2 k} )},
\label{cooltime}
\end{equation}

\noindent (White \& Frenk 1991), where $\Omega _{\mathrm b}$ is the
baryon fraction, $\Lambda (T)$ is the cooling function of the gas,
$f_{\mathrm g}$ is the fraction of the initial baryon density
remaining in gaseous form (which we normally assume to be 1),
${\mathrm G}$ is the gravitational constant, ${\mathrm m_p}$ is the
mass of a proton, $\mu$ is the mean molecular weight of the gas,
$V_{\mathrm c}$ is the circular velocity of the halo and ${\mathrm k}$
is Boltzmann's constant.

The evolution of the cooling radius determines the mass of gas cooling
per unit time.

\begin{equation}
\dot{M}_{\mathrm{cool}} (r_{\mathrm{cool}}) = {1 \over 2} {f_{\mathrm{g}} \Omega _{\mathrm{b}} \over \Omega _0} 
{V_{\mathrm{c}}^2(r_{\mathrm{cool}}) r_{\mathrm{cool}} (V_{\mathrm{c,max}},z) \over \mathrm{G} \mathit{t}_{\mathrm{0}} }.
\end{equation}

This can be combined with equation (\ref{eq:appLX}) to estimate the
total X-ray luminosity of the halo.

\subsection{Cooling in an NFW Halo}

\label{NFWcool}

NFW provide a fitting formula for the density profiles of dark matter
haloes in N-body simulations of hierarchical clustering.  This profile
is characterised by two parameters: $\delta _c$, a characteristic
density, and $r_{\mathrm{s}}$, a characteristic radial scale in the
halo. NFW find that $\delta _c$ and $r_{\mathrm{s}}$ are correlated in
any given cosmology so that the NFW profile provides a universal,
one-parameter model of the halo.  This formula provides a good
approximation to the density profile of haloes of all masses over at
least two orders of magnitude in radius and its shape is independent
of the values of the cosmological parameters.

Appropriate parameters for the galaxy haloes in any given cosmology
are determined by requiring that the measured circular velocity of the
galaxy corresponds to the maximum in the NFW rotation curve. This must
be fitted in an iterative fashion since the ratio between
$V_{\mathrm{c,vir}}$ and $V_{\mathrm{c,max}}$ itself depends on $c =
r_{200} / r_{\mathrm{s}}$.

Using the circular velocity profile of an NFW halo, the cooling radius
is defined by the relation

\begin{eqnarray}
t_0 & = & {192 \pi \over 49} {1 \over f_{\mathrm{g}}} {\Omega _0 \over \Omega _{\mathrm{b}}} {\mathrm{G m_p^2} \mathit{r}_{\mathrm{s}}^{\mathrm{2}} \over \mathrm{\Lambda} (\mathit{T}_{\mathrm{vir}})} {x_{\mathrm{cool}} \over x_{\mathrm{vir}}} \left( 1 +
cx_{\mathrm{cool}} \right) ^2 \nonumber \\ & & \times \left[ \ln (1 +cx_{\mathrm{vir}}) -
cx_{\mathrm{vir}}/(1+cx_{\mathrm{vir}}) \right] ^{-1}.
\end{eqnarray}

\noindent where $x = r / r_{200}$. We then obtain, 

\begin{eqnarray}
\label{eq:NFWLX}
L_{\mathrm{X}} & = & \dot{M}_{\mathrm{cool}} (r_{\mathrm{cool}}) V_{\mathrm{c}}^2(r_{200}) \nonumber \\ & &
\times \int _{x_{\mathrm{optical}}}^{x_{\mathrm{cool}}} x^{-2} {\ln (1 + cx) - cx / (1 + cx)
\over \ln (1 + c) - c / (1 + c)} {\rm d}x,
\end{eqnarray}

\noindent where $x_{\mathrm{cool}} = r_{\mathrm{cool}} / r_{200}$ and
$x_{\mathrm{optical}} = r_{\mathrm{optical}} / r_{200}$. This equation
must be integrated numerically for any given $c$.

\end{document}